# All-Chalcogenide Programmable All-Optical Deep Neural Networks

*Ting Yu[2,†], Xiaoxuan Ma[1,†], Ernest Pastor[3], Jonathan K. George[1], Simon Wall[3,4], Mario Miscuglio[1], Robert E. Simpson[2], Volker J. Sorger[1,*]*
\**sorger@gwu.edu*

[1]*Deptartment of Electrical and Computer Engineering, George Washington University, Washington DC, DC, USA*
[2]*Singapore University of Technology and Design, 8 Somapah Road, 487372 Singapore, Singapore*
[3]*ICFO - Institut de Ciencies Fotoniques, The Barcelona Institute of Science and Technology, Av. Carl Friedrich Gauss 3, 08860 Castelldefels (Barcelona), Spain*
[4] *Department of Physics and Astronomy, Aarhus University, Ny Munkegade 120, 8000 Aarhus C, Denmark*
[†]*These authors contributed equally to this work*

**Abstract –** Deeplearning algorithms are revolutionising many aspects of modern life. Typically, they are implemented in CMOS-based hardware with severely limited memory access times and inefficient data-routing. All-optical neural networks without any electro-optic conversions could alleviate these shortcomings. However, an all-optical nonlinear activation function, which is a vital building block for optical neural networks, needs to be developed efficiently on-chip. Here, we introduce and demonstrate both optical synapse weighting and all-optical nonlinear thresholding using two different effects in a chalcogenide material photonic platform. We show how the structural phase transitions in a wide-bandgap phase-change material enables storing the neural network weights via non-volatile photonic memory, whilst resonant bond destabilisation is used as a nonlinear activation threshold without changing the material. These two different transitions within chalcogenides enable programmable neural networks with near-zero static power consumption once trained, in addition to picosecond delays performing inference tasks not limited by wire charging that limit electrical circuits; for instance, we show that nanosecond-order weight programming and near-instantaneous weight updates enable accurate inference tasks within 20 picoseconds in a 3-layer all-optical neural network. Optical neural networks that bypass electro-optic conversion altogether hold promise for network-edge machine learning applications where decision-making in real-time are critical, such as for autonomous vehicles or navigation systems such as signal pre-processing of LIDAR systems.

## Introduction

Machine learning has become an essential part of societies, where systems are progressively exploiting (un)supervised deeplearning to detect objects, recognize speech, translate languages, or make decisions. However, demand for machine learning computing outweighs supply due to three bottlenecked requirements for information processing; (i) data movement across distributed networks or circuits, (ii) efficient performance of multiply-accumulate (MAC) operations enabling tensor operations such as matrix-vector multiplications (MVMs), and (iii) providing nonlinearity at each neuron, known as thresholding. Thus, next-generation efficient and high-performance artificial intelligence (AI) hardware must deliver each function, (i) to (iii), as optimized physical hardware since a one-fits-all approach leads to processor limitations. Such hardware optimization is known as algorithm-to-hardware homomorphism, and such synergistic co-design was performed in developing the deep all-optical photonic neural network introduced here. Currently, deeplearning algorithms are implemented in digital electronic van Neumann-type computing systems (i.e. systolic arrays[1] or SIMD units [2]). State-of-the-art graphic processing units (GPU) and tensor processing units (TPU) accelerators show performance values around 0.5-1 pJ/MAC, 0.5-1 TMAC/s/mm$^2$, 0.5-1 GMVM/s, and 1-2 μs delay[3]. While digital electronic AI hardware is able to provide seamless nonlinearity and memory functionality to store trained weights, there are two significant limitations: (a) data communication bottleneck where the signaling energy consumptions scales with the circuit capacitance, and b) algebraic operations (e.g. MAC, MVM, tensor) requiring power hungry and delay-prone electronic circuitry. Thus, machinery to coordinate the data movement involved in both weights and activations result in high overhead and delay. In fact for the last decade already, data movement outweighs data operations, opening a gap between these two domains. Optical signals are fundamentally well suited for communication, given the non-existence of an optical capacitance for traveling-wave systems. In addition to utilizing efficient photonic communication, here we discuss and introduce optical algebra both for linear but more importantly for nonlinearity, with the latter being critical for deep neural network architectures. Such depth is important since it is the key for many machine learning algorithms, especially those relying on gradient descent back-propagation during training.

Indeed, light is an established communication medium and is used to successfully address data movement (e.g. waveguides, fibers) and high fan-out (e.g. beam splitters). Thus, optics can seamlessly provide the required interconnectivity set by the distributed nature of neural networks. Furthermore, the compactness of photonic integrated circuits (PIC) paired with non-circuit-size related light propagation speeds (unlike electronic circuits), allows for picosecond-short signal delays between neurons. Despite such prospects, the more interesting question is, whether one can perform mathematical operations, namely linear MVM and nonlinear thresholding, efficiently and fast in the optical domain? The question is fair, since typical compute-functionality such as nonlinearity, memory, or gain, are cumbersome to be performed optically and/or on-chip. Yet, linear algebra, such as for MAC operations and MVMs are implementable in photonics enabled by either interference effects or by passing an optical signal through an optical-index modulating linear device (e.g. modulator[4]). Such PIC-based linear operation processors include Mach–Zehnder interferometer[5], ring-filters[6], or all-optically implemented by nonlinear light-matter interactions, for example. Conceptually, optical linear information processing can be further parallelized by exploiting time[7], space[8,9], or the wide spectrum such as in wavelength division multiplexing schemes[6]. However, in the context of chip-based photonic AI technologies, implementation options are limited, and interferometric-approaches are challenging to control due to high sensitivity and do not scale well (e.g. $N^2$ MZI's are required to implement $N$ neurons). Most limiting, yet, is the conversion to electronics after each MVM operation[5]; optically accelerating MVMs just to be limited by electronic circuits for the nonlinear thresholding operation, followed by iteratively re-using the same photonic chip, hurts system delay, and makes such photonic neural network approaches questionable.

Here, we introduce a novel all-optical photonic neural network paradigm harnessing the full potential from light-based machine learning acceleration featuring; (i) no domain crossings (i.e. no O-E-O conversions needed); (ii) linear and nonlinear operations performed optically on-chip and via the same material class hence reducing fabrication complexity and future system cost; (iii) photonic nonvolatile memory to store the neural network weights offering near-zero static power consumption; (iv) cascadability by providing nonlinear activation functions (NLAF) optically on-chip enabling deep architectures; (v) response times for inference tasks of 10's ps enabling near real-time decision making; and (vi) demonstrating the above in silicon photonic platform where the functional light-matter-interactions are enhanced by heterogeneous integration by a single foundry-near material class (i.e. chalcogenides). Overall, this paradigm for photonic neural networks is attractive because the computation and memory takes place in the same physical location thus bypassing the infamous memory-access bottleneck and hence suppressing latency; i.e. *in-memory photonic computing*.

Phase change materials (PCM) are leading candidates for setting synaptic weights in optical neural networks. This is because they exhibit substantial optical contrast between structurally ordered and disordered states, are foundry processable, and commercially proven in related technologies, such as re-writeable optical discs, and more recently in non-volatile electronic memory. The most commonly employed PCM is $Ge_2Sb_2Te_5$ (GST), which is capable of fast reversible transitions in its refractive index. Its potential as an electro-refractive memory element in integrated photonic chips has been demonstrated in coupler-based[10,11], cavity-based[12–14] and absorptive schemes[15], and GST memory elements can be programmed optically (all-photonic[16]) or electro-thermally by either passing a current through the GST such as using micro-heaters[17,18]. However, GST's application for in-memory photonic computing is hindered by (1) it high optical absorption from inter-band transitions at photon energies above 0.7 eV[19], which limits the depth (i.e. multi-layer) of the neural network; (2) hence this loss also restricts the number of optical levels (bit-density)[16,20] which impacts the accuracy of inference tasks; and (3) the phase transition time which limits the latency (signal delay) through the neural network.

To address the inefficiencies of current photonic neural networks, we adopt reprogrammable photonic devices made of two different chalcogenides. To overcome the high optical absorption of GST, we adopt $Sb_2S_3$ to set the synaptic weights. $Sb_2S_3$ is a wide-bandgap PCM and is therefore transparent across the telecoms spectral bands whilst retaining an almost unity change in refractive index between its amorphous and crystalline states. In contrast to phase transitions, we employ an underexploited highly nonlinear disruption of resonant bonding in crystalline GST for the NLAF. Combining optical nonlinear spectroscopy measurements with full-wave and interconnect simulations[21], we show an all-optical artificial neuron not requiring any O-E or E-O conversions until the very last layer of a deep architecture. This concept synergistically embeds the synaptic nonvolatile (programmed) weights and volatile NLAF with the same material (here GST) framework heterogeneously integrated atop a silicon photonics neural network platform. The photonic neuron exploits nonvolatile structural transition (Δn = unity) while exhibiting ultra-low-losses (κ = $10^{-4}$ at NIR) PCM[22] for programmable MVM operations, and experimentally demonstrates volatile rapid transitions[23] as NLAF. Exemplary, the quantized synapses are obtained through a cascade of ultra-low-losses $Sb_2S_3$-based[22] broadband 4-bit switches operating as an optical FPGA and here referred to as photonic programmable phase-change array ($P^3A$), while the NLAF is achieved by deploying a reversible and rapid (< 10 ps) non-thermal modulation of the dielectric function of a GST film. Here, we reveal for the first time, that the co-participation of two competing phenomena in the GST material, i.e. fs-laser induced depletion of electrons from resonantly bonded states and near-infrared free-carrier optical absorption, produce a strong nonlinearity, whose shape ultimately allows for an accurate prediction during inference phase (**Fig. 1a**). A multi-level (2-layer) perceptron feed-forward neural network based on these novel all-optical photonic neurons, is trained using a photonic hardware model[24] to accurately simulate the complete process and learn the weights including analog noise. The network is ultimately emulated on an open-source machine learning framework. After training, classification is demonstrated on hand-written digits with a high level of accuracy (>90%) despite the relatively high levels of variability in the NLAF and weights.

The herein introduced all-optical deep neural network paradigm enables noise-robust intelligent in-memory computing with potential sub-picosecond latency for inference, which is limited only by the time-of-flight of the photon and the rapid thermalisation of the GST, but not by phase-transitions. Assuming optical input signals, the network can perform classification of the MNIST handwritten digits consuming just 0.1μJ of optical energy, which is needed for the nonlinear activation modules.

## Results

The building block of our network is an artificial photonic neuron (**Fig. 1b**). Analogous to a biological neuron, this artificial neuron acquires optical inputs and performs a weighted addition (linear MAC operation) using the photonic programmable phase-change array ($P^3A$) consisting of a network of programmable photonic latches to control 800 nm signals (SiN waveguide platform). The synaptic weights, which are acquired by off-line training, are then programmed as refractive index changes in an ultra-low loss $Sb_2S_3$–based multistate photonic phase change memory material[22]. These $Sb_2S_3$ memory elements are embedded in asymmetric directional couplers and is programmed using electrical Joule heat or all-optically using a pump laser[15,16,25] (**Fig. 1c**). The optical output of this $P^3A$ is amplified in the optical domain using either on-chip semiconductor optical amplifiers (SOAs) or Erbium Doped Fibre Amplifiers (EDFAs). All-optical amplification may be required for large multi-layer networks to ensure sufficient signal-to-noise ratio (SNR) at the back-end photodetector during high signal-speeds (e.g. 50-100 Gbps). After the linear weighted addition the optical signals interact with the GST layer nonlinearly by destabilizing resonant bonds, altering its complex permittivity of the GST film on a picosecond time scale[23,26]. This GST, now being in an excited state, influences the transmission of a second laser signal at 1550 nm, which is either attenuated or fed to a second layer of a photonic neural network, thus enabling a rapid NLAF (**Fig. 1d**).

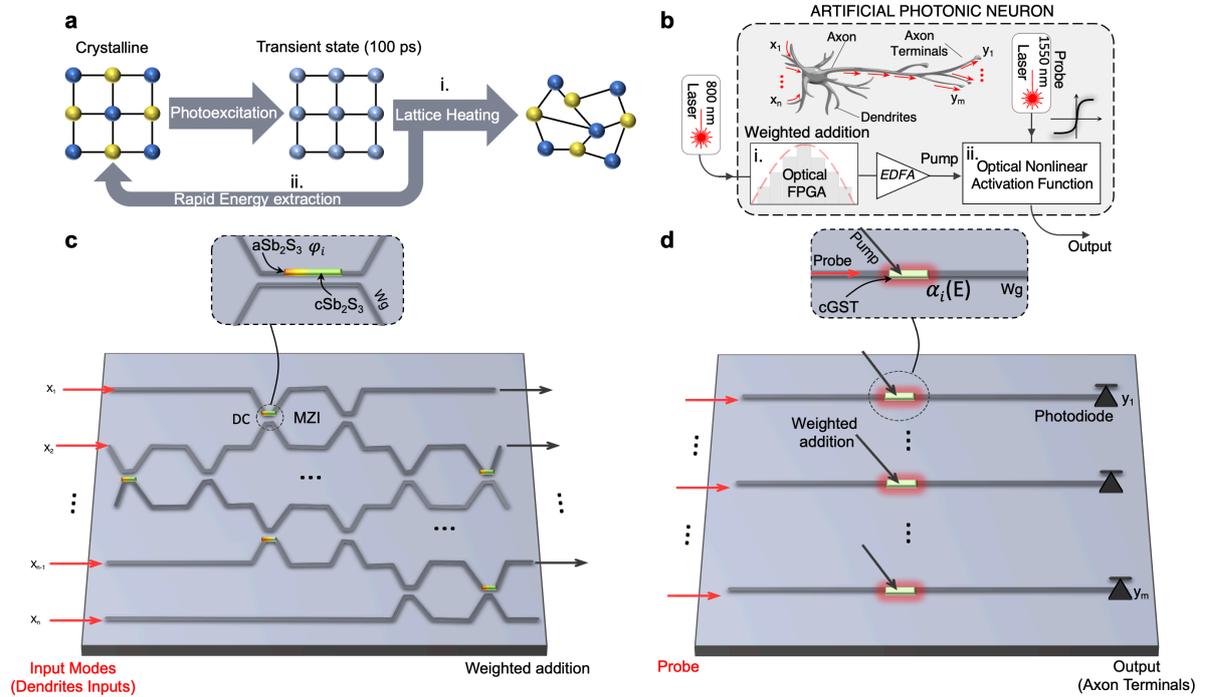

**Figure 1. a, An artificial photonic neuron based on structural and electronic transitions in phase change materials.** Ultrafast photoexcitation forms a transient state by removing the resonant bonds and changing the optical properties. Rapid energy extraction (using silicon slab waveguide as heat-sink) prevents amorphization allowing for a volatile transition. Successive heating, after several picosecond, causes lattice heating, which thermally melts the GST. **b,** Schematic of the artificial photonic neuron which comprises an photonic programmable phase-change array (P$^3$A), programmed according to quantized synaptic weights, and all-optical nonlinear activation function (NLAF) based on phase change materials (PCM). Inputs at 800 nm laser source are linearly combined according to a cascade of programmable switches (SiN-based photonics platform), the resulting optical power is amplified, e.g. using erbium doped fiber amplifier (EDFA), and used as pump signal for the optical NLAF module which uses a 1550 nm probe signal. The output of the NLAF is injected in the Si-based photonic layer. **c,** Weighting mechanism and summation rely on a cascade of $Sb_2S_3$-SiN hybrid photonic switches which performs the same operation as an optical equivalent of an FPGA (Field Programmable Gate Array). The PCM is placed on one of side of the directional coupler (DC), and according to the portion of material written (crystalline to amorphous) the light is partially configured to the cross state. The phase of the film is thermally set, with a resolution of ~1 µm yielding to a total of 16 distinguishable states (4-bit). Fully amorphous PCM film configure light in a cross state of the DC switch, while fully crystalline results in a bar state. The hybrid configurations in between bard- and cross states represent the intermediate states. **d,** The NLAF module consists of a single mode hybrid silicon waveguide covered for 1 µm by a 30 nm thick $Ge_2Sb_2Te_5$ layer. A 1550 nm TM polarized light is used for sensing the nonlinear variation of the effective refractive index induced by the fs-laser coupling into the PIC e.g., from free space. Plasmonic antennas could be used for enhancing light matter interaction. After the NLAF the signal could be either detected using integrated photodetector or passed to a second stage photonic waveguide.

# Programmable photonic phase-change array (P³A) for weighted addition

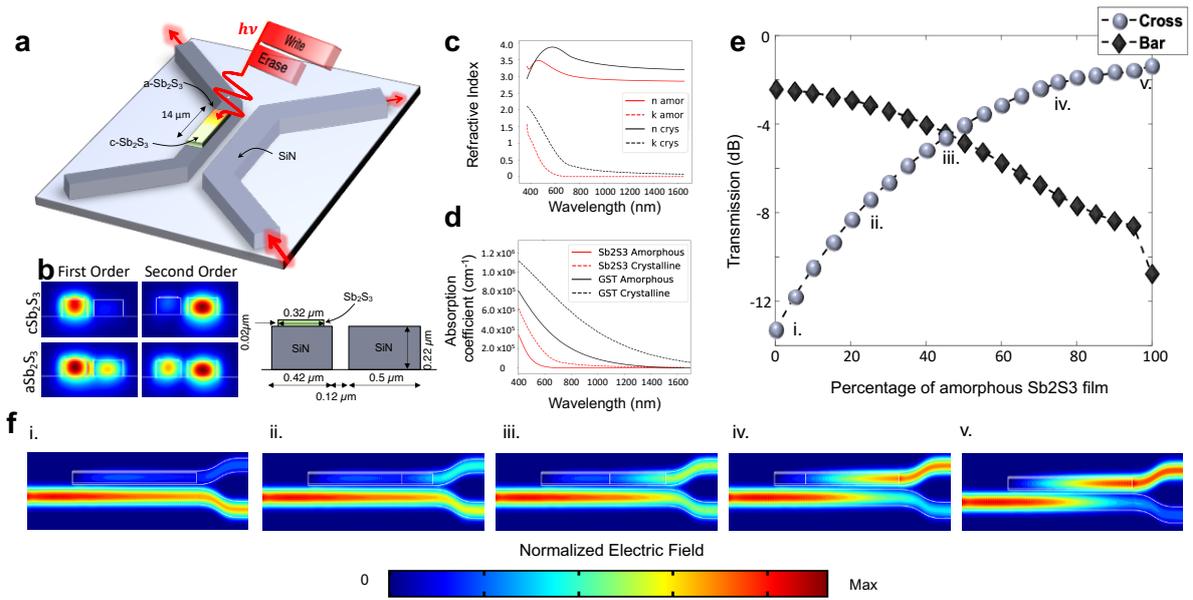

**Figure 2. Optical-equivalent of an electro-thermally controlled Programmable Photonic Phase-change Array (P³A) based on photonic nonvolatile memory on-chip. a)** Schematic representation of the silicon nitride asymmetrical directional coupler (DC) operating at 800 nm. The DC is composed of two waveguides, a bare silicon nitride waveguide (SW, 500 nm width) and a hybrid $Sb_2S_3$-on-silicon nitride hybrid waveguide (HW, 420 nm width), which are separated by a 120 nm gap. The thickness of the $Sb_2S_3$ is 20 nm. A free space laser source can be used for writing portion of the $Sb_2S_3$ film deposited on top of the hybrid waveguide with a writing resolution of 1 μm. According to the percentage of the portion of the $Sb_2S_3$ layer that is amorphized, the percentage of light which switches to the HW changes proportionally. Numerical approach used Comsol Multiphysics.

The core of the P³A is an asymmetrical electro-optic programmable MZI-based switch operating at 800 nm (**Fig. 1c**). The asymmetric coupling region comprises a bare silicon nitride waveguide (SW) and a hybrid $Sb_2S_3$-on-silicon nitride hybrid waveguide (HW), as illustrated in **Fig. 2a**. $Sb_2S_3$ is selected for a low absorption coefficient at 800 nm compared to GST (**Fig. 2 c-d**) which yields to devices characterized by lower insertion losses in this frequency range. Optimization of the dimension of the switch ensure smallest footprint, length = 14 μm for 4-bit resolution, which is relatively compact compared to recent MZI-based weighting approaches[5] (**Fig. 2e**). The MZI switch is designed to transmit via the SW arm of the coupler (bar-state) with low insertion loss (2 dB) when $Sb_2S_3$ is fully crystallized (**Fig. 2e.i** and **Fig.2f.i**). However, as the $Sb_2S_3$ is progressively amorphized, the SW light progressively couples towards the HW and the transmittance of the switch in the cross-state following the expected cosine squared function (**Fig. S2**) of the portion of amorphized layer (**Fig. 2e.i-v** and **Fig.2f.i-v**). Thus, when the HW low-loss $Sb_2S_3$ layer is fully amorphized, the switch meets the phase-matching condition for transverse electric (TE) polarization, leading to an insertion loss (IL) of 1.6 dB at 800 nm (**Fig. 2f.v** and **Fig.2e.v**). An analog or multilevel coupler is then formed by controlling the length of the crystallised $Sb_2S_3$ film. We find that 4-bit quantization provides a reasonable trade-off between writing resolution for programming the synaptic weights (~1μm) and an inference accuracy[27]. Note, higher resolutions are possible too, just linearly increasing chip footprint. The required PCM-length is determined by the downstream detector and the signal clock-speed to ensure a SNR > 1 is provided, and a sufficiently high bit-error-rate (BER) for the choses machine learning application. Further details of the P³A performance are presented in the SI.

## Photonic nonlinear activation function (NLAF)

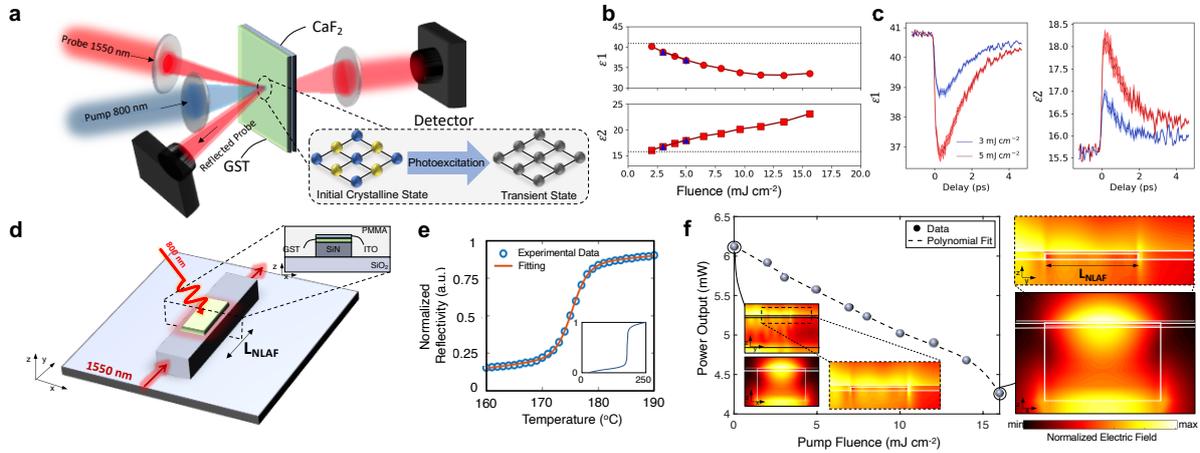

**Figure 3. Nonlinear activation function (NLAF) of photonic neurons. a,** Pump-probe physics and temporal response of the all-optical nonlinear activation function based on resonant bond-state depopulation in GST thin films. Schematic of the pump probe setup. The 800 nm pump bears the weighted-addition (MAC) signal of 4-bit resolution from the upstream photonic perceptron (**Fig. 1&2**). Pump: fs-laser 35 fs pulses, $\lambda$ = 800 nm. Probe: $\lambda$ = 1550 nm. **b,** Peak change in dielectric function after photoexcitation for different pump fluences. Blue triangles represent repeat-measurements after raising the fluence to over 17 mJcm$^{-2}$, demonstrating no permanent change is induced in the sample. **c,** Time traces of the dielectric function showing that the majority of the changes recover in the first few picoseconds as the resonant bonding state reforms, enabling for rapid volatile NLAF functionality on-chip at near-zero delay when performing inference tasks. The shaded area indicates the change in extracted dielectric function assuming a range of initial thicknesses for the GST layer, from 24-30 nm. **d,** Nonlinear activation function (NLAF) based on an electro-absorptive pump and probe scheme The electronic transition in the PCM, heterogeneously integrated on top a SiN provides the all-optical nonlinearity. The pump signal, i.e. the P³A outputs modulated according to the previous weights, is coupled from free-space, amplified and focused onto each GST segments of the NLAF modules. **e,** Transmittance of the 1550 nm probe signal as function of the pump-signal fluence (result of the weighted addition of the optical interference unit). Insets show the normalized electric field distribution for a propagating hybrid TM mode in the xz and yz planes when the dielectric constant of the GST film is (i.) in the crystalline phase and (ii.) has electronically transitioned due to the depletion of resonant bonds induced by the fs pump laser. **f,** Crystallisation of GST. The change in optical reflectivity with temperature shows a logistic function is differentiated to determine the crystallisation temperature (see Methods and SOM). Inset: full temperature range, from room temperature to 250 °C.

Next, we experimentally test the on-chip all-optical NLAF via nonlinear time-resolved spectroscopy to determine the temporal response, power efficiency, and modulation depth of the GST-based photonic NLAF (**Fig. 3**). The transient changes to the near-normal reflectivity and transmissivity (**Fig. 3a**) are simultaneously recorded by a $\lambda$ = 1500 nm time-delayed probe pulse with a pump at $\lambda$ = 800 nm delivering the 4-bit weighted MAC signal from the upstream photonic perceptron. The dielectric function is then obtained by numerically inverting the Fresnel equations for the multi-layer structure assuming all changes occurred in the c-GST film (see Methods). The maximum change of the complex dielectric function occurs 0.75 ps after the pump pulse (**Fig. 3b**) enables a rapid volatile NLAF functionality at near-zero delay in performing inference tasks. Regarding the underlying physical effect providing this observed nonlinearity, the change in the signal results from the combined effects of a Drude-like effect due to free carrier generation and the loss of resonant bonding, which primarily effects the real part of the dielectric function[23]. The transient response at two typical fluences displays a rapidly recovering response, within several picoseconds after excitation (**Fig. 3c**). A smaller offset persists for hundreds of picoseconds corresponding to the thermal change.[19]

For the all-optical deep neural network, we design a NLAF photonic module that consists of a hybrid Si waveguide upon with a 1 μm long, 30 nm thick strip of crystalline GST patterned (**Fig. 3d**). This short device design can modulate incoming signals at $\lambda = 1550$ nm by 1.5 dB, which shows that extremely compact photonic NLAF components are possible. To achieve these results, we first used a finite-difference time-domain approach to simulate the effect of $\lambda = 800$ nm pump light on $\lambda = 1550$ nm signal transmission. The pump signal arriving from the linear weighting module ($P^3A$, **Fig. 1c**) and the amplification stage, depopulates the resonantly bonded states in the crystalline GST and causes a concomitant decrease in the effective refractive index of the guided $\lambda = 1550$ nm TM mode. The higher the pump fluence, the lower the overall power signal output because the excitation increases the number of free carriers thus increasing absorption. The 3.5 dB insertion loss is caused by the non-negligible extinction coefficient of the c-GST layer in the crystalline state (**Fig. 3e**). A Cavity-based modulation scheme could be engineered to compensate for the higher insertion losses, but this would reduce the spectral bandwidth of this nonlinear device. The energy consumption of a single nonlinear node is 0.1 nJ, and hence quite low. This compares well with bulkier resonant absorbers and electro-optic NLAF modules, since here the device has a footprint of just 1μm and does not require additional circuitry, which is critical for compact neural network architectures. Although, non-linearity could be further enhanced using plasmonics[15], it would be at the expense of increased insertion losses due to the ohmic losses in the plasmonic metal.

**Training and Performance**

To test the performance of the all-chalcogenide all-optical deep neural networks, we explore the performance of a 3-layer fully connected neural network composed of the photonic-$P^3A$ and photonic NLAF modules using Googles TensorFlow (**Fig. 4**). The system was trained on the MNIST dataset, which is a well-known standard that comprises of 60,000 grayscale images of handwritten digits. The first layer receives the inputs from the images of the handwritten images and does not apply any weighting to the inputs. The second layer has 100 neurons, which receive inputs from the first layer with an all-to-all connection (100x100) and performs weighting (nonlinear quantized during inference). The all-optical NLAF[28] are placed between two consecutive layers on each input connection (**Fig. 4a-b**). The third layer reduces the dimensionality of the network and comprises just 10 neurons, which represent the outputs 0 to 9. The network is trained on the MNIST dataset utilizing both photonic $P^3A$ and NLAF modules. Unlike digital electronics where noise is artificially added to the training data to increase the networks robustness, photonic neural networks being analog systems bear a degree of intrinsic noise. Testing the robustness of this all-photonic deep neural network, we monitor the inference accuracy as a function of switching variability (i.e. noise) from both the P$^3$A and the NLAF (**Fig. 4c**)[24]; the general trends are as expected, namely a higher accuracy for less noise. The highest accuracies are achieved when 0.01% noise is added to the maximum signal swing at the neuron's output. A possible explanation is that deliberately adding training noise can fine-tune the network for inference operation on physical noisy input signals. However, if the training noise is beyond 0.1% of the signal strength, the inference accuracy decreases, as expected. Strategies for further increasing the accuracy of the network include: 1) increasing the bit resolution of the P$^3$A and increasing the both number of neurons in each layer and the network depth; 2) adopt advanced quantization algorithms to better represent the information during training, and furthermore; 3) gradually add the quantization constraints in a training-retraining flow to converge to a 'better' local optimum during the training.

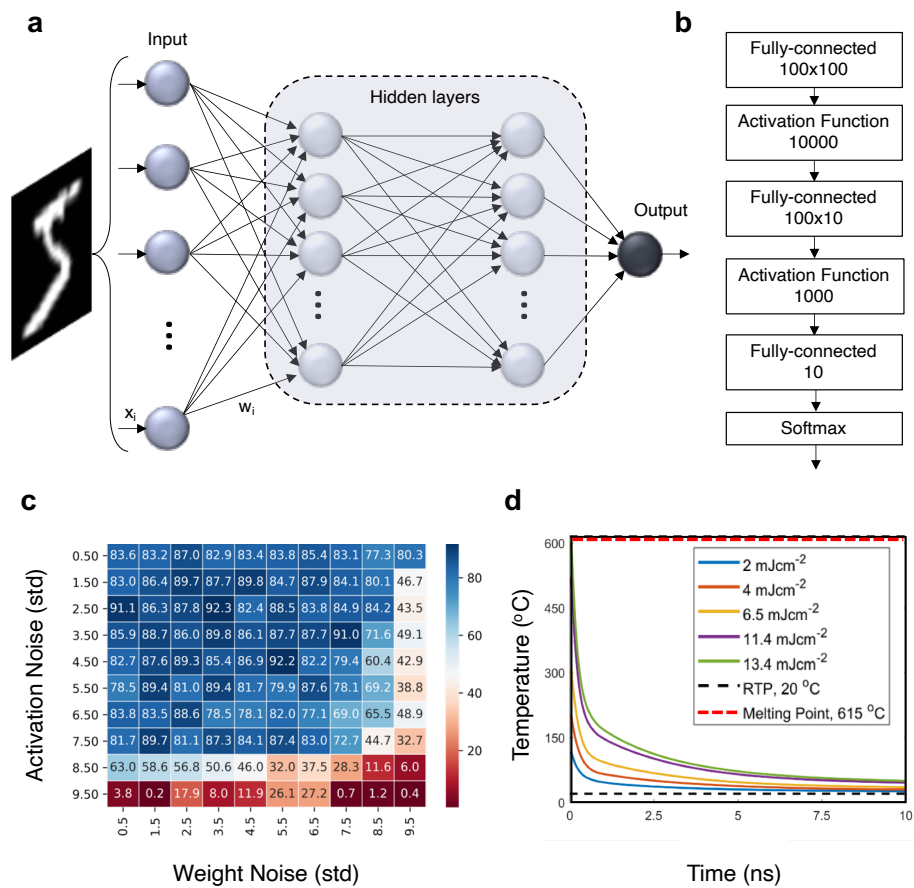

**Figure 4. All-Photonic Deep Neural Network Performance. a,** Schematic of the fully connected network composed by 2 layers of 100 neurons. **b,** Dataflow graph of the NN which comprises 100 neuros and 2 fully connected layers. **c,** Accuracy results for the inference on unseen data for NN trained with 2% of Gaussian noise. The evaluation of the effect of NLAF module and synaptic quantized weights on inference accuracy. Soft max operation is considered to be performed electronically. **d,** The speed limit of this all-photonic deep neural network depends on the thermal transients of GST films providing the neuron's nonlinearity, (assuming a fixed and programmed kernel. The temperature values are extracted from literature[23] and the cooling behavior was simulated using the finite element method on COMSOL Multiphysics.

Next, we comment on the overall performance of this all-photonic deep neural network paradigm; as previously reported, the lattice temperature achieved for below-threshold excitation (below amorphization) increases linearly with excitation fluence (up to 15 mJ cm$^{-2}$), therefore the GST material patch must be quenched efficiently to enable high repetition rates. With higher fluences approaching GST's melting point, the device must be able to efficiently dissipate heat (see SOM). The current device setup which has the Si$_3$N$_4$ waveguide can be further optimized to facilitate heat dissipation (cooling) via adding a metal sheets (yet there is a risk to increase insertion loss) or alternatively adding a graphene layer, which has a high thermal conductivity[29].

Assuming a fixed kernel (i.e. the linear MVM is programmed into the PCM materials), then the temporal bandwidth (speed) of the neural network performing passive inference tasks is dictated by the thermal response time of the GST element in the NLAF module to cool down. The underlying physical effect of the nonlinearity is based on electric-field induced depopulation of p-orbital electrons in the resonantly bonded GST crystal. However, as these electrons thermalize, the GST will increase in temperature. If the frequency of the pump pulses, corresponding to the data rate of the MVM signals, is too high, the GST temperature will equilibrate at temperatures above the 950 K melting temperature. The molten properties of GST are very different to the crystalline state[30], and therefore it is important to ensure the pump laser repetition rate does not cause the GST element to melt. A thermal-transient analysis shows a MHz range for the threshold pump pulse frequency for stable operations for all tested fluences (**Fig. 4d**). The lattice temperatures were obtained from previous literature, where it was reported that the lattice temperature achieved for below-threshold excitation (below amorphization) increases linearly with excitation fluence (up to 15 mJ cm$^{-2}$). The maximum operating frequency of the laser pulse was approximated to be 2.4 ns (~400 MHz) using 1-D heat diffusion equation (see SOM).

This operating speed can be extended further by optimizing the thermal design of the nonlinearmodule, coherently pumping the A1 mode of the GeTe$_4$ units in the GST crystal[31], and using interfacial phase change materials, which have a larger thermal conductivity[32]. Moreover, adding a highly thermal conductive material like graphene[29] in the device would dissipate the heat more efficiently. Ultimately, the network's theoretical inference latency is limited by the time-of-flight through the network, which is in the ps-range, and for the network design discussed here, the neural network's latency is 20 ps. Larger networks will have a linear-path-length longer latency, whilst the standard deviation in pulse arrival times will increase with the number of neurons due to the greater number of routes that photons can take through the network. Nonetheless, the intrinsic speed of the NLAF result in latencies at least three orders of magnitude lower than equivalently size electrical solid-state neural network implantations.

**Discussion**

Emerging neural network accelerators should be benchmarked against exciting solutions. Comparing the scaling vectors of an electronic accelerator (e.g. GPU[3]) with recent developments of photonic-electronic-hybrid neural networks[5] to the all-photonic deep neural network discussed in this work, we find the following insights (**Table 1**): (1) the performance gains from an electronic accelerator stem from the fact that they do not require any domain crossings (e.g. optic-to-electronic-to-optic, OEO; or digital to analog) and are able to operate at a high bit precision. Its power- and delay limitation are typically determined by communication and memory access bottlenecks. (2) Photonic-electronic hybrid systems (e.g. Ref [[5]]), on the other hand, offer rapid optical MAC operation performing the linear MVM with a lower bit resolution (given by the DAC/ADC), but are limited by electronic-to-photonic crossings to perform the NLAF in electrical circuits. Hence, the delay gain from PICs is traded in for performing the required NLAF of deep neural networks. Thus, implementations only process only a single a layer of the neural network at a time (or portions of that layer depending on the input vector length and available photonic hardware) using the available photonic accelerator iteratively, thus sacrificing run-time. Power consumption wise, these hybrids are limited by the laser, the OEO conversions, and to program the weights into the photonic chip requiring both DACs and drivers for the electro-optic components (e.g. modulators or phase-shifters). (3) All-optical neural networks, in contrast, preserve the networks' depth and keep the to-be-processed input signal in the optical domain across multiple layers for neural network, until the final layer where the OE conversion occurs using high-speed photodetectors. The bit-resolution is not limited by DACs but by setting the memory bit-level, while power consumption is constrained by the optical NLAF. Since updates for the weights of the NN are available only rarely (i.e. training is time consuming and new training data sets are often a safe-guarded trade-secret, such as the 'valuable' customer data), storing the weights in nonvolatile memory is logical, thus preserving power. That is, consider the alternative; the weights are realized in photonic NNs with electro-optic (EO) components (phase shifters, or modulator-like devices). Since these photonic EO components do no retain their state (volatile), one has to read the weight state from a memory on an on-going basis. Such data-fetching from an off-chip memory (DRAM), or even from an on-chip SRAM is rather time and power costly, and requires use of very power-hungry DACs. A solution for this challenge is to deploy the PCMs as nonvolatile memory, thus enabling for near zero-static power consumption, once the weights are programmed. Note, that a bit-resolution of 4-6 is sufficient to perform inference tasks adequately and preserves power, a strategy that is also pursued in electronic machine learning accelerators such as the NVIDIA T4 system, for example. From a fabrication and hence system cost point-of-view, the ability to utilize PCM-based materials for both the linear and nonlinearoptical algebra allows for reduced complexity and offering synergies. Considering that our neural network comprises 1100 activation function modules, the energy consumption of the entire network for performing inference would be just 0.1μJ. All the activation functions can be triggered at unison using a passive laser beam spatially patterned, using a digital micromirror array, according the network topology. However, an experimental implementation of the entire network would require the fabrication of a relatively large number of neurons, which might require repeaters for boosting the optical signal.

| Hardware | Memory | NLAF | NN Layers | Domain Crossings? | Bit Resolution | Power Limitation | Delay Limitation | Ref |
|---|---|---|---|---|---|---|---|---|
| GPU | Centralized | E | $M$ | None | Full bit Precision (32 bits) | Communication with Memory + Operations | Communication with Memory (ms) + Digital NLAF (> 0.1 ns) | [3] |
| Photonic-Electronic Hybrid NN | Centralized + DAC + MZI mesh network | E | 1 | Digital ⇔ Analog, & Photonic ⇔ Electronic | DAC limited (e.g. 8 bits) | Laser + DAC + OEO | 100 GHz PD + electric NLAF (digital, > 0.1 ns) + Communication /w Memory (ms) | [5] |
| All-photonic DNN | On-chip (compute in memory) | AO | $M$ | Photonic ⇔ Electronic* | Levels of memory (4-6 bits) | NLAF + Laser | 100 GHz PD + 0.1ps for Optical NLAF | This work |

**Table 1.** Conceptual system comparison of photonic, electronic-photonic hybrid neural network accelerators to electronic systems (i.e. GPU). An all-optical (AO) photonic chip-integrated neural network (NN) offers to realize deep (DNN) multi-layer architectures avoiding optic-to-electronic-to-optic (OEO) conversion until the very last layer performed by a photodetector (PD). Key is to realize an efficient nonlinear activation function (NLAF) at each layer without OEO delays or signal losses. In contrast, electronic-photonic hybrid systems perform the multiply algebra photonically, such as via a Mach Zehnder Interferometer (MZI) mesh network, and do require an OEO conversion to add nonlinearity electronically at each layer, which adds temporal delay and system overhead, thus limits scaling. Since updating weights from NNs is a relatively rare event, photonic machine learning accelerators should make use of PIC-integrated photonic nonvolatile memories, rather than relying on off chip memory, which introduces both delay and power overheads such as from digital to analog converters (DAC).

We have demonstrated through a combination of simulations and fs transient optical spectroscopy that all-chalcogenide all-photonic deep neural networks can be noise robust and capable of performing intelligent tasks with high accuracy. The key to the network accuracy and speed is the ability to momentarily destabilize the highly polarizable p-orbital electrons which are delocalized in certain chalcogenide crystals, such as GST. This results in radically nonlinear changes to the material's refractive index. A second important material used in this work is the recent report wide bandgap phase change material, $Sb_2S_3$, used here program the neuron's weights exemplary with 4-bit resolution. When these effects are combined in a photonics on-chip all-photonic neural network architecture, the result is a highly accurate system that can be used to identify patterns in optical signals at 10-100 Gbps data rates. This architecture is the foundation on which all-optical photonic neural networks can be developed to break the memory-compute dichotomy by circumventing unnecessary electro-optic conversions, and ultimately allowing to implement deeper neural networks for performing more complex inference tasks.

From a practical perspective, femtosecond lasers tend to be expensive and bulky. Therefore, we foresee that all-optical network implementations find applications in data centres where space and power is less important and optical interconnects are commonplace. However, desktop machine learning application specific photonic neural network accelerator chips will require compact laser pump sources, such as diode lasers capable of producing hundred picoseconds optical pulses. However, the electric field generated by such laser pulses is orders of magnitude too low to depopulate resonant bonds in GST. However, many crystalline antimony-based materials exhibit nanosecond duration highly nonlinear changes to their optical constants when heated with nanosecond laser pulses[33]. Indeed, a related effect has also been reported for electrical Joule heating GST-based electrical phase change memory cells[34].. We conclude, therefore, that all-optical deep neural networks can be designed and implemented in photonic integrated circuits by performing optical algebra for both the linear matrix vector multiplication operation but also for the nonlinearity required at each layer of the neural network. Summarizing, the main advantages of all-optical photonic NNs is to allow for deep (i.e. multi-layer) architectures, short temporal delays just limited by the back-end photodetector, and the weights can be stored in photonic memories while 'passively' performing the multiply algebra. The only active power consumed is related to the external laser source for triggering the volatile transition of the GST film of the NL modules, which are amortized across the high data rate (e.g. 50-100 Gbps). The herein introduced all-optical deep neural network paradigm enables noise-robust intelligent in-memory computing with potential sub-picosecond latency for inference, which is limited only by the time-of-flight of the photon and the rapid thermalisation of the GST, but not by phase-transitions. Assuming optical input signals, the network can perform classification of the MNIST handwritten digits consuming just 0.1μJ of optical energy, which is needed for the nonlinear activation modules. Optical neural networks that bypass electro-optic conversion altogether hold promise for network-edge machine learning applications where decision-making in real-time are critical, such as for autonomous vehicles, or navigation systems including LIDAR, but also as machine learning accelerators.

## Materials and Methods

*Waveguide Modelling and Design.* An eigenmode solver within Comsol Multiphysics was used to find the widths of the HW waveguide, which is tuned to obtain the same effective refractive index as that of the single-mode SW when the $Sb_2S_3$ is in the amorphous state, so the phase matching condition (cross-state) is satisfied for the 100% amorphous layer (**Fig. S1**). In this configuration, the HW width is 420 nm, which is 80 nm thinner than the monomodal SW. A coupling length was of $\frac{\lambda_0}{2}(n_{\text{eff},Sb_2S_3-1} - n_{\text{eff},Sb_2S_3-2})$~14μm was found to be optimal, where $n_{\text{eff},Sb_2Se_3-1,2}$ are the effective refractive indices of the super-modes of the 2-waveguide system (odd and even) at $\lambda_0 = 800$ nm. The gap between the waveguides found to be 120 nm (**Fig. 2b**).

*Sputtering.* A 50.8 mm diameter 99.9% pure GST target was used to sputter the GST films in an Argon environment using the AJA ORION 5 sputtering system. The RF power was 30 W and the deposition pressure was 0.5 Pa, which resulted in a deposition rate of 0.093 Å/s. The deposition rate was calibrated using step profilometry of films deposited for a fixed time.

*Crystallisation.* The crystalline GST (c-GST) film was prepared in a two-step process; amorphous GST was first deposited on a quartz (SiO$_2$) substrate using radio frequency magnetron sputtering. The sample was then annealed at 183 °C for 30 minutes in a 4 sccm flowing argon atmosphere. This induced a structural phase transition from amorphous to the cubic crystalline state. The phase transition temperature is observed as an abrupt change in optical reflectivity when the material is heated.[35] Details of the sputtering conditions and how the crystallization temperature was determined can be found in the methods section. To determine the phase transition temperature, we recorded the change in optical reflectivity as we heat the material. The as-deposited amorphous GST thin film was heated to 250 °C, at a heating rate of 5 °C/min, and with an Ar gas flow rate of 4 sccm. The material reflectivity was recorded each time the heating stage temperature increase by 1 °C. To ensure consistency, this sample was from the same sputtering batch as the sample in **Fig. 3** of the main text. The inset of **Fig. 3f** shows the change in optical reflectivity as the temperature increases; with the main figure focusing on the region where the abrupt change in optical reflectivity occurred. This abrupt change corresponds to the structural phase transition from amorphous to face-centred cubic (FCC).[35]

We then differentiate the fitted reflectivity curve in **Fig. 3f**. The maximum point of the differential curve is the phase transition temperature. The fitted reflectivity curve, shown in equation (1), was used to simplify the differentiation operation. Least- squares fitting was used to determine the coefficients of (1). From **Fig. 3f**, we observe that the equation fits well with the data points. This was also reflected in the goodness of fit indicator, $R^2$, which gave a value of 0.9998. The coefficient values of equation (1) can be found in Table 1 of supporting information.

$$\text{Normalized Reflectivity} = \sum_{n=1}^{2} \frac{a_n}{1 + b_n e^{(-c_n (T - d_n))}} \quad (1)$$

where $a$, $b$, $c$ and $d$ are constants determined through a least-squares fitting algorithm, while $T$ represents the heating stage temperature. The coefficients do not bear any physical meaning and is solely used to simplify the differentiation operation. The differential of figure M1 (a) is shown in **Fig. S1** in the supplementary material. We observe that the phase transition temperature is approximately 176 °C. To ensure that the material is fully crystallised, we heat the sample at 183 °C for 30 minutes.

**Time-resolved experiments and retrieval of the dielectric function:**
The transient optical measurements were performed using a commercial regenerative 5KHz Ti:Saphire amplifier (Coherent) that generated the pump pulse (800 nm) and pumped a commercial OPA (Light conversion TOPAS) to generate the probe pulse (1500 nm). A mechanical chopper at 500Hz was used to modulate the pump and the changes in reflectance *ΔR/R (t)* and in transmittance *ΔT/T (t)* were simultaneously recorded using two InGaAs detectors. The changes in the optical constant (refractive index *n* and dielectric function) were estimated using a procedure previously reported[36]. In brief, the changes in transmission *T(n',d, λ)/T(n$_o$,d,λ)-1* and reflection *R(n',d, λ)/R(n$_o$, d, λ)-1* for the multilayer structure (GST/SiO$_2$ in our case), were calculated using a transfer matrix method at a given wavelength λ, thickness d. Multiple reflections in the SiO$_2$ substrate were ignored. n$_o$=6.5+1.2i and n' are the initial and perturbed complex refractive indices respectively. These value of n' that gives the best fit to both data sets simultaneously is found at each delay. The effects of small variations in the GST thicknesses (24,27 and 30 nm respectively) were found to only give a small error in the retrieved refractive index. In order to ensure a uniform excitation area, the pump spot (180x100 um) was significantly bigger than the probe.